\documentclass[%
 reprint,
 amsmath,amssymb,
 aps,
]{revtex4-2}
 \usepackage{natbib}

\usepackage{hyperref}
\hypersetup{
    colorlinks=true,
    linkcolor=blue,
    citecolor=blue,
    urlcolor=blue
}

\newcommand{\beq}{\begin{equation}}
\newcommand{\beqn}{\begin{eqnarray}}
\newcommand{\eeq}{\end{equation}}
\newcommand{\eeqn}{\end{eqnarray}}

\usepackage{amsmath,amsfonts,amssymb}

\usepackage{appendix}
\usepackage{graphicx}
\usepackage{dcolumn}
\usepackage{bm,color}

\begin{document}

\title{ Electrostatic Doping of Moiré Superlattices Controls the Optical Fingerprint of a $WSe_2$/Twisted Bilayer Graphene heterostructure}

\author{H. E. Hannachi}
\email{hannachi.houssemedine@fsb.ucar.tn}

 \affiliation{Laboratoire de Physique des Mat\'eriaux, Facult\'e des Sciences de Bizerte, Universit\'e de Carthage, 7021 Zarzouna, Tunisie.}

\author{ S. Jaziri}
 \email{Sihem.Jaziri@fsb.rnu.tn}
  \affiliation{Laboratoire de Physique des Mat\'eriaux, Facult\'e des Sciences de Bizerte, Universit\'e de Carthage, 7021 Zarzouna, Tunisie.}

\date{\today}

\begin{abstract}
We theoretically investigate the optical response of the $WSe_2$ monolayer vertically stacked on twisted bilayer graphene (tBG) under electrostatic doping. In this heterostructure, the doped moiré superlattice of tBG generates a spatially modulated electrostatic potential that couples to the electron and hole constituents of the $WSe_2$ exciton through the screened Coulomb interaction, thereby providing a microscopic mechanism for exciton coupling to the charge-density landscape of the moiré Fermi sea. Focusing on twist angles below the magic angle (1.1°), where electronic correlations are strongly enhanced, we demonstrate that electrostatic doping induces pronounced modifications of the excitonic absorption spectrum, including the emergence of multiple satellite resonances. We show that these spectral features originate from the interplay between the exciton center-of-mass and internal degrees of freedom, together with doping-induced hybridization between Rydberg exciton states of different orbital symmetry. The calculated spectra are in good agreement with previously reported experimental observations and theoretical predictions. Our results establish electrostatic doping in tBG as an effective means of engineering excitonic states in adjacent semiconducting monolayers and provide a pathway toward novel optical sensor devices.
\end{abstract}
\maketitle

\section{Introduction}
Two-dimensional (2D) semiconductors, particularly monolayers of transition metal dichalcogenides (TMDs), have emerged as an outstanding platform for exploring many-body phenomena in condensed matter physics owing to their remarkable optoelectronic properties and rich excitonic landscape. Excitons dominate the optical response of TMD monolayers because the reduced dielectric screening and strong quantum confinement give rise to exceptionally strong Coulomb interactions between electrons and holes. Consequently, these materials exhibit large exciton binding energies, typically on the order of several hundred meV \cite{Chernikov2014}, which can be efficiently tuned by external electric and magnetic fields \cite{Molas2017,Zhu2023,Stier2016,Kamban2019,Hannachi2024}, the dielectric environment \cite{Itzhak2025}, or mechanical strain \cite{Kumar2024}.
The coupled spin and valley degrees of freedom in TMDs monolayers provide an additional avenue for engineering excitonic states. The valley-selective optical selection rules enable the resonant excitation of bright intra-valley excitons at the inequivalent K and K' valleys using circularly polarized light, offering unprecedented opportunities for valleytronic and quantum-optical applications \cite{Xiao2012}.
\\

Similar to the hydrogen atom, the exciton internal motion  is quantized into a discrete Rydberg series consisting of states such as 1s, 2s, and $2p_{\pm}$. The s-type exciton states are optically bright and can be directly accessed through one-photon absorption. In contrast, the p-type Rydberg states are optically dark in linear spectroscopy because of the corresponding selection rules and therefore require nonlinear optical techniques, such as two-photon spectroscopy \cite{Pedersen2021}, for their experimental observation.
\\
\begin{figure*}
    \centering
    \includegraphics[width=0.46\linewidth]{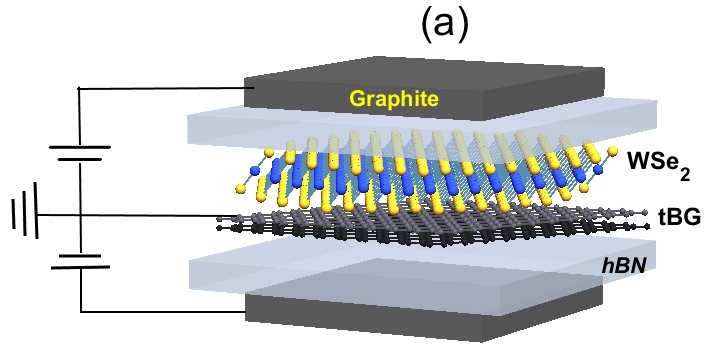}
    \quad
    \includegraphics[width=0.41\linewidth]
    {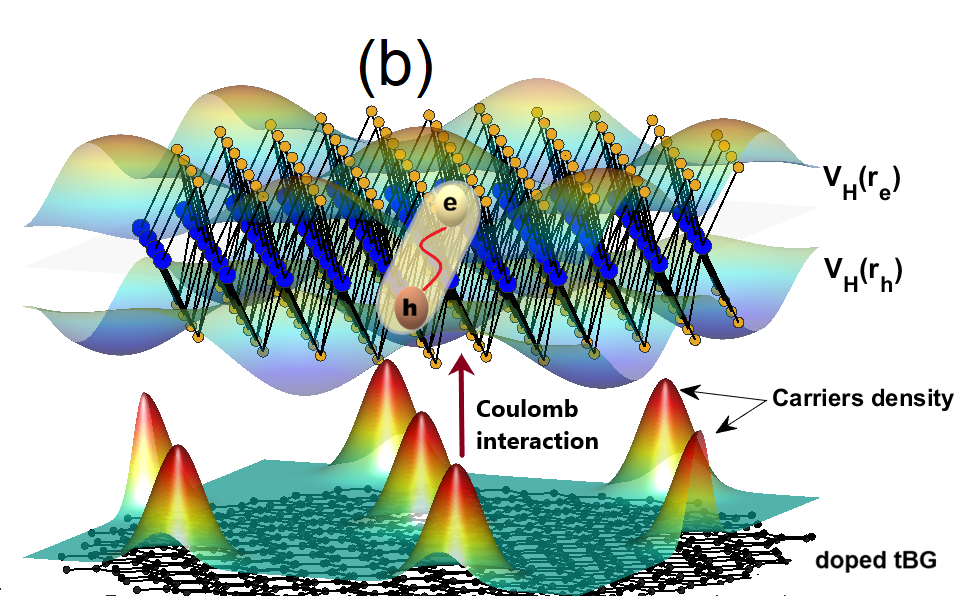}
    
    \caption{(a) Schematic illustration of the studied $hBN/WSe_2/tBG/hBN$ heterostructure. The $WSe_2$ monolayer is positioned adjacent to twisted bilayer graphene, which is electrostatically doped to control the carrier density in the moiré superlattice. (b) Cartoon illustrative of the mechanism by which the carrier density in the doped moiré tBG generates a spatially periodic electrostatic potential in the $WSe_2$ monolayer. This  potential acts on the electron and hole constituents of the exciton through the respective potentials $V_H(\mathbf{r}_e)$ and $V_H(\mathbf{r}_h)$. }
    \label{fig1}
\end{figure*}

More recently, excitonic Rydberg states in moiré superlattices-- formed by vertically stacking 2D materials with a finite twist angle and/or lattice mismatch-- have attracted considerable attention because the emergent moiré potential and interlayer degree of freedom profoundly modify the  optical response. The long-moiré period gives rise to spatial confinement of excitons, leading to the formation of moiré exciton states with tunable electronic and optical properties.

An especially active research direction concerns electrostatically doped moiré superlattices, in which the injected carriers form a moiré Fermi sea \cite{Arsenault,Hu2025,Lu2024,He2024,Hu2023}. The interplay between the periodic moiré potential  and strong Coulomb interactions gives rise to a rich variety of correlated electronic phases. Depending on the moiré filling factor, these systems can host generalized Wigner crystal states \cite{Xu2020,Kim2026,Kim2025,You2026}, Mott insulating phases \cite{Amine2025,Lian2024,Lian2023,Qi2026},  fractional Chern insulators \cite{Perea2025}, and other exotic many-body phenomena like the exciton-polaron formation \cite{Pichler2026,Evrard2025}. Consequently, doped moiré superlattices have emerged as a highly tunable platform for investigating strongly correlated physics in 2D materials.
\\

In the present paper, we theoretically study an electrostatically doped
tBG system placed adjacent to a monolayer of
$WSe_2$. Motivated by recent experimental observations~\cite{Hu2023} and theoretical
predictions~\cite{Lu2024}, which have demonstrated that carrier doping can profoundly
modify the optical response of moiré excitons and enhance the visibility of
excitonic Rydberg states, we focus on the emergence of multiple satellite
resonances in the optical spectra. Such features have been observed for twist
angles below the magic angle and are particularly pronounced in the range of
$0.6^\circ$--$0.8^\circ$. Their emergence is expected to originate from the
interplay between the internal and center-of-mass degrees of freedom of the
exciton, together with the hybridization of excitonic Rydberg states with
different symmetries, such as the $ns$ and $2p_{\pm}$ states.
In this context, we develop a detailed microscopic model of the underlying
$WSe_2$/tBG heterostructure and investigate the dependence of its optical
response on electrostatic carrier injection. In particular, we analyze the
evolution of the moiré excitonic Rydberg spectrum as a function of experimentally
tunable parameters, including the  twist angle and the interlayer separation
between the tBG and $WSe_2$ monolayer. 
\\

 The paper is organized as follows. In Sec. \ref{sec2}, we introduce the theoretical framework used to describe the electrostatically doped hBN encapsulated $WSe_2/tBG$  heterostructure, including the excitonic Hamiltonian and the interaction with the doped moiré fermi sea. In Sec.~\ref{sec3}, we present our main results, focusing in particular on the
dependence of the optical absorption spectrum on carrier filling, while explicitly accounting for the interlayer degrees of freedom of tBG. Finally, Sec. \ref{sec4} summarizes our main conclusions.

\section{Theoretical model} \label{sec2}

 We consider a bright intra-valley exciton formed in a  $WSe_2$ monolayer placed adjacent to twisted bilayer graphene (tBG), where the latter is electrostatically doped  [see Fig.~\ref{fig1}(a)]. The carrier redistribution within the doped tBG moiré superlattice generates a spatially modulated electrostatic potential that acts on the electron and hole constituents of the exciton through the screened Coulomb interaction. This effective periodic potential reflects the interaction between the exciton and the charge density landscape of the doped moiré Fermi sea, as shown in [Fig.~\ref{fig1}(b)]. As a consequence, the exciton experiences a filling-dependent moiré potential inherited from the underlying tBG superlattice.
Within this framework, the moiré exciton can be described by the following two-body Hamiltonian

\begin{equation}\label{eq1}
    H_X=Eg+\frac{\textbf{p}_e^2}{2m_e}+\frac{\textbf{p}_h^2}{2m_h}+U_{sc}(|\textbf{r}_e-\textbf{r}_h|)+V_H(\textbf{r}_e)-V_H(\textbf{r}_h).
\end{equation}
  
Here, $E_g$ is the $WSe_2$ band-gap energy, $\mathbf{p}_{\alpha}$ and $m_{\alpha}$ denote the momentum and effective mass of the charge carrier, respectively, with $\alpha=e,h$ labeling the electron and hole.  The screened electron--hole Coulomb interaction described by the Fourier Rytova--Keldysh potential \cite{Rytova,keldysh}
\\

\begin{equation}\label{Rytova}
    U_{\mathrm{sc}}(|\mathbf{r}_e-\mathbf{r}_h|)=\int U_{sc}(q)e^{i\textbf{q}(\textbf{r}_e-\textbf{r}_h)}d^2q/(2\pi)^2,
\end{equation}

in term of q-space screened Coulomb potential $U_{sc}(q)=2\pi e^2/(\epsilon q 
(1+
\ell_0q))$, with $\epsilon$ denotes the effective dielectric constant of the surrounding dielectric environment, while
$\ell_0=2\pi\chi_{\mathrm{2D}}/\epsilon$
is the screening length of the $WSe_2$ monolayer, with  $\chi_{\mathrm{2D}}$ denoting its 2D polarizability. 
We also define $\textbf{r}_\alpha$ as the charge carrier position vector. 

The last two terms [Eq.~(\ref{eq1})] account for the effective periodic electrostatic potential generated by the doped tBG moiré superlattice. Owing to the opposite charges of the electron and hole, the potential enters the Hamiltonian with opposite signs, ensuring the charge neutrality of the exciton. This filling-dependent electrostatic modulation reflects the coupling between the exciton and the charge density distribution of the moiré Fermi sea. The periodic potential may be written as

\begin{equation}
V_H(r_\alpha)=\sum_{\textbf
{G}_j}U_H(G_j;\nu)e^{i\textbf{G}_j.\textbf{r}_{\alpha}},  
\end{equation}

 where $
\mathbf{G}_j=
\frac{4\pi}{\sqrt{3},a_m}
\left[
\cos\left(\frac{(j-1)\pi}{3}\right),
\sin\left(\frac{(j-1)\pi}{3}\right)
\right],$ with $j=1,\ldots,6$,
denoting the six first-shell moiré reciprocal lattice vectors of the mini-Brillouin zone (mBZ). The moiré period is approximately given by $
a_m \simeq a_0/2\sin{(\theta_t/2)},
$
where $a_0=2.46$ A° is the Graphene lattice constant and $\theta_t$ is the interlayer twist angle.
The Fourier component of the effective electrostatic potential is given by 

\begin{equation}
U_H(G_j;\nu)=U(G_j)\,\rho(G_j; \nu), 
\end{equation}

where  $U(G_j)$ is the momentum-space interlayer screened Coulomb interaction, reflecting the interaction between  $WSe_2$ exciton and tBG moiré Fermi sea. it can be written as \cite{Lu2024,Danovich}

\begin{equation}
  U(G_j)=\frac{2\pi e^2 e^{-G_jd}}{\epsilon G_j[(1+G_j\ell_{s_1})(1+G_j\ell_{s_2})-\ell_{s_1}\ell_{s_2}G_j^2 e^{-2G_jd}]},  
\end{equation}
\\

here, $\ell_{s_1}$ and $\ell_{s_2}$ denote the screening lengths of the $WSe_2$ monolayer and the tBG, respectively.  We take $\ell_{s_1}=4/\epsilon$ nm, consistent with the screening length used for the electron--hole Coulomb interaction in [Eq~(\ref{Rytova})]. Since the charge-carrier screening length in tBG is expected to be substantially larger than that in  the  $WSe_2$ monolayer, we take  $\ell_{s_2}=28/\epsilon$ nm.
The parameter $d$ represents the vertical separation between the $WSe_2$ layer and the tBG. In our calculation, we set $d=0.6$ $\mathrm{nm}$ and $\epsilon=4$,  corresponding to an hBN-encapsulated heterostructure.
\\

The Fourier component of the carrier density modulation in the doped tBG moiré superlattice, is modeled by the Gaussian form

\begin{equation}
  \rho(G_j;\nu)
\simeq
\nu \rho_m
e^{-(G_j\sigma_\nu)^2/4},  
\end{equation}
 
where $\nu$ is the moiré filling factor and $\rho_m=2/(\sqrt{3}a_m^2)$ denotes the carrier density per moiré unit cell. This form reflects the accumulation of doped carriers around the AA regions of the tBG moiré superlattice. We assume that the carrier distribution associated with each AA site can be approximated by a Gaussian wave packet characterized by a width $\sigma_\nu$. Owing to the combined effects of the Pauli exclusion principle and Coulomb repulsion, the spatial extent of the charge distribution is expected to increase with carrier filling. Following the work~\cite{Lu2024}, we therefore adopt the phenomenological parametrization
$
\sigma_\nu \simeq 2\left(1+0.1|\nu|\right)\mathrm{nm}.
$, which provides an adequate description for the calculations presented below. The Gaussian form of $\rho(\mathbf{G}_j;\nu)$ captures the dominant long-wavelength features of the charge redistribution within the doped tBG moiré superlattice while preserving the underlying triangular symmetry.

We adopt the [Eq.~(\ref{eq1})]  in terms of the internal and COM degrees of freedom, the exciton Hamiltonian then takes the form

\begin{equation}\label{eq.5}
    H_X=E_g+\frac{\textbf{p}^2}{2\mu}+U_{RK}(r)+\frac{\hbar^2\textbf{Q}^2}{2M_X}+V_H(\textbf{R},\textbf{r}),
\end{equation}

 Here, $
\mathbf{p}=-i\hbar\nabla_{\mathbf r} $ denotes the relative momentum operator associated with the electron--hole relative coordinate
$
\mathbf r=\mathbf r_e-\mathbf r_h.
$ The reduced and total exciton masses are defined as
$
\mu=\frac{m_em_h}{M_X}
$ and $
M_X=m_e+m_h$,  respectively. The exciton COM momentum operator is given by
$
\mathbf Q=-i\nabla_{\mathbf R},
$  where
$
\mathbf R=
\frac{m_e\mathbf r_e+m_h\mathbf r_h}{M_X} $ is the COM coordinate.  The last term in [Eq.~(\ref{eq.5})] corresponds to the effective exciton electrostatic potential induced by the doped tBG superlattice. Owing to its dependence on both the relative and COM coordinates, this potential couples the internal and COM degrees of freedom of the exciton. Consequently, the periodic potential can induce simultaneous transitions between different excitonic Rydberg states and COM Bloch states. Explicitly, the interaction term can be written as

\begin{equation}
    V_H(\textbf{R},\textbf{r})=\sum_{\textbf{G}_j} U(G_j)\rho(G_j;\nu)e^{i\textbf{G}_j.\textbf{R}}F_{\textbf{G}_j}(\textbf{r}),
\end{equation}
where   
  $F_{G_j}(\textbf{r})=e^{i\alpha_e \textbf{G}_j.\textbf{r}}-e^{-i\alpha_h \textbf{G}_j.\textbf{r}}$ defines the exciton form factor,
in terms of $\alpha_{e(h)}= m_{h(e)}/M_X$. We note that the second an third  terms in [Eq.~(\ref{eq.5})] describe the exciton internal motion, which determines the excitonic Rydberg series, whereas the last two terms govern the COM motion, including the doping-induced periodic mean-field potential.
\\

To solve the Schrödinger equation for the moiré exciton, one may consider the momentum scattering by the reciprocal-space moiré vectors
$
\mathbf g_j=l_1\mathbf G_1+l_2\mathbf G_2,
$
where $l_1$ and $l_2$ are integers.  These reciprocal lattice vectors mediate momentum transfer within the mBZ and enter the moiré-exciton wavefunction through the Bloch-like form
\\

 \begin{center}
   $
\Psi_X(\mathbf{r},\mathbf{R})
=\langle \mathbf{r},\mathbf{R}|n,\ell;\mathbf{Q}+\mathbf{g}_j\rangle
=\phi_{n,\ell}(\mathbf{r})
e^{-i(\mathbf{Q}+\mathbf{g}_j)\cdot\mathbf{R}}.
$  
 \end{center}

Here, $n$ and $\ell=\pm(n-1)$ denote the principal and angular momentum quantum numbers, respectively, which characterize the excitonic Rydberg orbitals, namely the $s$ ($\ell=0$), $p_\pm$ ($\ell=\pm1$), and $d\pm$ ($\ell=\pm2$) states. The internal exciton wave function is expressed in the 2D hydrogenic basis as 
$
\phi_{n,\ell}(\mathbf{r}) = e^{i\ell\theta}R_{n,\ell}(r)
$\cite{Hannachi2025},   
where $\theta$ is the internal angular angle and  $R_{n,\ell}(r)$ is the radial wave function.
\\

After diagonalization of the internal motion Hamiltonian, we denote the resulting internal exciton eigenstates by $|\lambda\rangle$. The basis states of the moiré exciton Hamiltonian are then constructed as
$|\lambda;\mathbf Q+\mathbf g_j\rangle$. Accordingly,  the matrix element Hamiltonian is written as

 \begin{multline}\label{Eq.H}
\langle \lambda';\textbf{Q}+\textbf{g}_j'|H_X|\lambda;\textbf{Q}+\textbf{g}_j\rangle=\\\left[
\frac{\hbar^2|\mathbf Q+\mathbf g_j|^2}{2M_X}
+E_\lambda^X
\right]
\delta_{\lambda\lambda'}
\delta_{\mathbf g_j,\mathbf g_{j'}}
\\
+\sum_{\mathbf G_j}
U_H( G_j;\nu)
I_{\lambda'\lambda}( G_j)
\delta_{\mathbf g_{j'}-\mathbf g_j,\mathbf G_j}
+\mathrm{H.c.}.
\end{multline}

Here,
$E_\lambda^X=E_g-E_\lambda^b$
is the bare exciton energy, with $E_\lambda^b$ denoting the binding energy of the internal exciton eigenstate $|\lambda\rangle$. Furthermore, we define the exciton kernel

\begin{center}
   $
I_{\lambda'\lambda}(\mathbf{G}_j)
=\int d^{2}r\,
\psi_{\lambda'}^{*}(\mathbf{r})
F_{\mathbf{G}_j}(\mathbf{r})
\psi_{\lambda}(\mathbf{r}),
$ 
\end{center}

with internal exciton eigenvectors 

\begin{center}
$\psi_{\lambda}(\textbf{r})=\sum_{n,\ell} D_{n\ell}^{(\lambda)}\phi_{n,\ell}(\textbf{r})$    
\end{center}

 and $D_{n\ell}^{(\lambda)}$ are the numerical eigenvector coefficients. By expressing the reciprocal lattice vector as
$
\mathbf{G}_j
=G_j(\cos\phi_G,\sin\phi_G),
$ with $\phi_G$ is the azimuthal phase  of the COM momentum, and employing the Jacobi--Anger expansion,
$
e^{ix\cos(\theta-\phi_G)}
=
\sum_{m=-\infty}^{\infty}
i^{m}
J_{m}(x)
e^{im(\theta-\phi_G)}.
$
Consequently, one obtains

\begin{equation}
     I_{\lambda'\lambda}(G_j)=\sum_{n'\ell'}\sum_{n\ell}D_{n'\ell'}^{(\lambda')*}D_{n\ell}^{(\lambda)}I_{n'\ell',n\ell}(G_j),
\end{equation}
with 
\begin{multline}\label{kernal}
    I_{n'\ell',n\ell}(G_j)=2\pi e^{i\delta \ell\phi_G} i^{\delta\ell}\int  R_{n'\ell'}(r)\bigl[ J_{\delta\ell}(\alpha_e G_j r)\\-(-1)^{\delta\ell}J_{\delta\ell}(\alpha_h G_j r)\bigr] R_{n,\ell}(r)rdr.
\end{multline}

 Here, $\delta\ell=\ell-\ell'$, and $J_m(x)$ denoting the Bessel function of the first kind of order m. The exciton kernel $I_{\lambda'\lambda}(\mathbf G_j)$ describes the coupling between the exciton internal motion and the COM   induced by the periodic electrostatic potential. Through the Jacobi--Anger expansion, the potential contains angular harmonics of all orders and can therefore couple Rydberg states with $\delta\ell=0,\pm1,\pm2,\ldots$, leading to a mixing of excitonic states with different orbital symmetries. However, in the special case $me=m_h$ ($\alpha_e=\alpha_h$), the matrix elements between states with identical orbital symmetry vanish identically ($I_{\lambda'\lambda}(G_j)=0$), and only transitions with odd parity $\delta \ell=\pm1,\pm3,\pm5...$, are allowed.
 \\
 
The coupling is constrained by the threefold rotational symmetry $C_3$ of the moiré lattice. The conserved quantum number is the total angular momentum modulo three, yielding the selection rule 
$\delta m_{\rm cm}+\delta\ell = 0 \pmod 3,$ 
where $\delta m_{cm}=m_{cm}'-m_{cm}$ in terms of   the COM angular momentum quantum number associated with the eigenvalue of the $C_3$ rotation operator,
$C_3|\Psi_{COM}\rangle=
e^{-i2\pi m_{\rm cm}/3}
|\Psi_{COM}\rangle ,$
with $m_{\rm cm}=0,\pm1$ labeling the three irreducible representations of the $C_3$  group symmetry. Consequently, only exciton states belonging to the same total $C_3$ symmetry sector can hybridize through the periodic potential.

\section{Results and discussion} \label{sec3}

\begin{figure*}
    \centering
    \includegraphics[width=.43\linewidth]{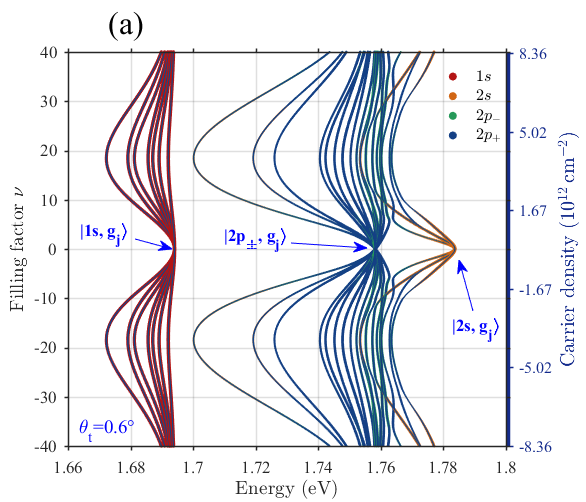}
    \quad
        \includegraphics[width=.36\linewidth]{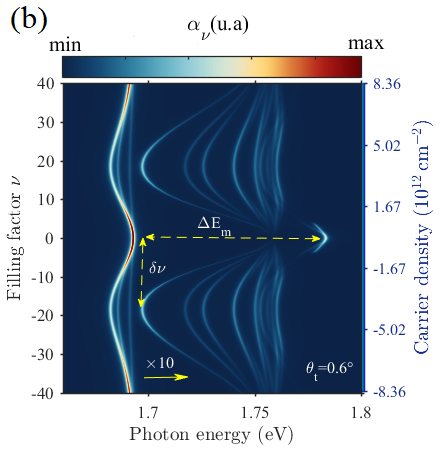}
    \\
    \includegraphics[width=0.4\linewidth]{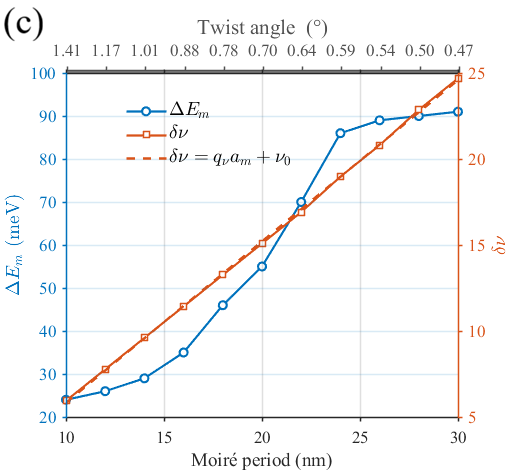}
    \quad
        \includegraphics[width=.4\linewidth]{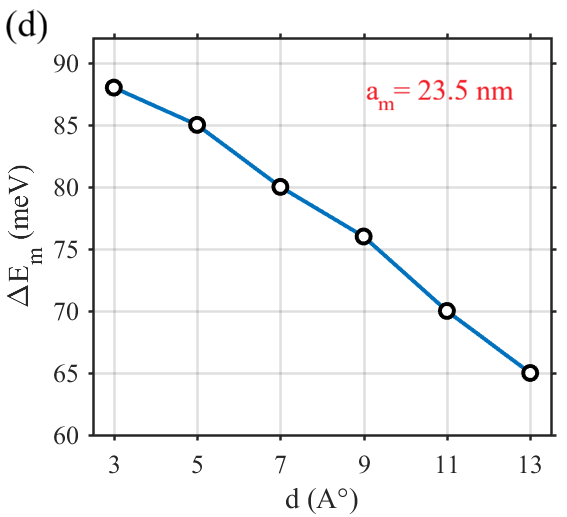}
    \caption{(a) Filling factor dependence of the bright moiré-exciton states in the $WSe_2$/tBG heterostructure at a twist angle of $\theta_t=0.6^\circ$ ($a_m=23.5$~nm), showing the states $|1s,\mathbf{g}_j\rangle$, $|2p_{\pm},\mathbf{g}_j\rangle$, and $|2s,\mathbf{g}_j\rangle$, with the corresponding projected Rydberg-state weights indicated by the marker size and color. (b) Corresponding absorption map as a function of filling factor (carrier-density). Here, the carrier-density axis is calculated as $\nu\times\rho_m$. The dashed yellow vertical and horizontal double arrow indicate the characteristic filling factor $\nu_m$ and energy scale  $\Delta E_m$, respectively. (c) Dependence of the  quantities $\Delta E_m$ and $\delta\nu$ on the moiré period (twist angle). (d) Dependence of $\Delta E_m$ on the interlayer separation d, at fixed moiré period $a_m=23.5$nm.}
    \label{fillingPL}
\end{figure*}

\begin{figure*}
    \centering \includegraphics[width=0.32\linewidth]{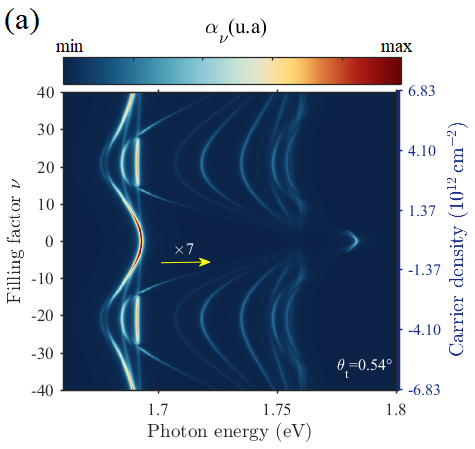}
    \quad
\includegraphics[width=0.3\linewidth]{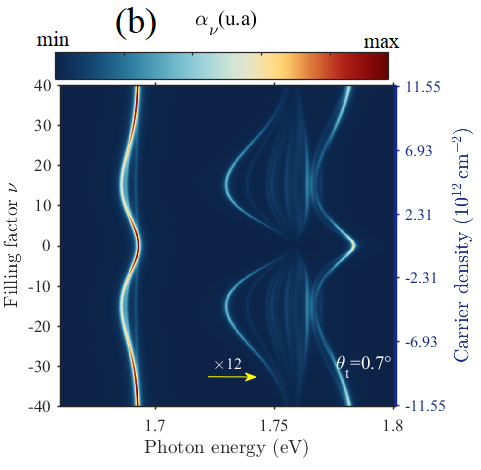}
    \quad
\includegraphics[width=0.32\linewidth]{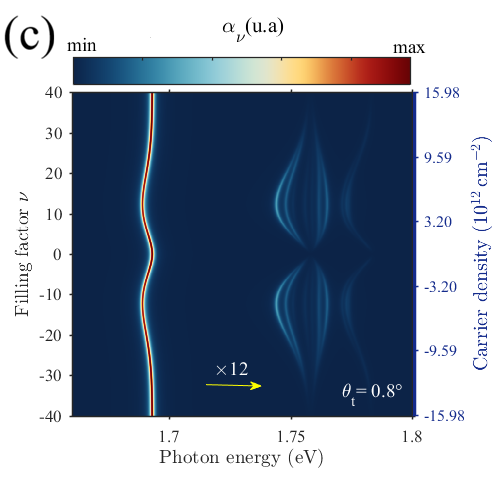}
    \quad
    \\
    \includegraphics[width=0.9\linewidth]{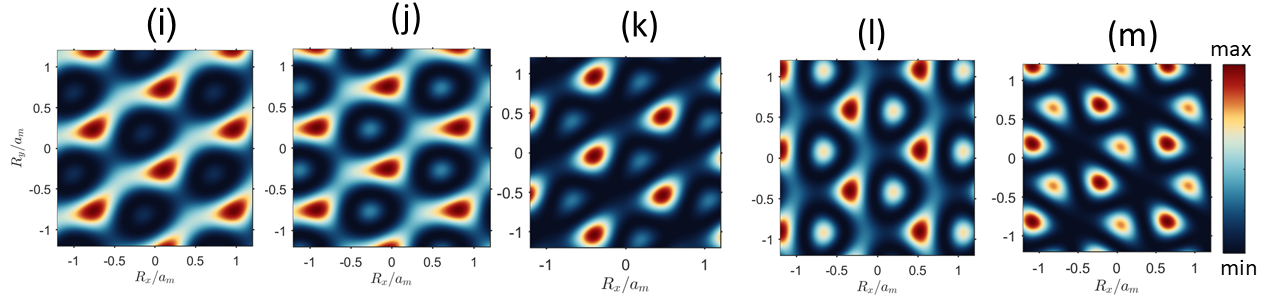}
    \caption{  (a--c) Filling-factor dependence of the absorption maps for twist angles of $\theta_t=0.54^\circ$, $0.7^\circ$, and $0.8^\circ$, respectively. (i--m) Corresponding real-space COM density distributions $ \left|\int \Psi(\mathbf{r}=0,\mathbf{R})d^2\mathbf{R}\right|_{\mathbf{Q}=0}^2$, for the 8th--12th bright states at a fixed twist angle of $\theta_t=0.45^\circ$ and filling factor $\nu=10$. }  
    \label{fillingPL2}
\end{figure*}

 In [Fig.~\ref{fillingPL}(a)], we present the dependence of the optically active moiré-exciton states (\textbf{Q}=0) on
 the filling factor $\nu$ (left axis), together with the corresponding carrier density (right axis), for a twist angle of $\theta_t=0.6^\circ$, corresponding to a moiré period of $a_m=23.5$~nm. The associated moiré-unit-cell density is $\rho_m\simeq 2.1\times10^{11}~\mathrm{cm}^{-2}$. The spectrum is resolved according to the projected orbital character of the exciton states, with the dominant contributions identified as the $1s$, $2p_{\pm}$, and $2s$ Rydberg states.
The exciton spectrum is obtained by numerical diagonalization of the Hamiltonian in [Eq.~(\ref{Eq.H})], yielding a total of $n^2N_{\mathrm{cm}}=252$ eigenstates. Here, the internal exciton basis is truncated at principal quantum number $n=6$, while $N_{\mathrm{cm}}=7$ accounts for the COM states associated with the first shell of reciprocal moiré-lattice vectors in the mBZ, namely, ${\mathbf{g}_j}={0,\pm\mathbf{G}_1,\pm\mathbf{G}_2,\pm\mathbf{G}_3}$. The resulting spectrum exhibits a pronounced non-monotonic dependence on the filling factor for the moiré-exciton states $|\lambda;\mathbf{g}_j\rangle$, where the relevant internal exciton states are represented by the Rydberg basis ${ |1s\rangle, |2p_+\rangle, |2p_-\rangle, |2s\rangle }$.
For the adopted electron and hole effective masses, $m_e=0.4m_0$ and $m_h=0.5m_0$ (in the unit of the free electron mass $m_0$), and a band-gap energy of $E_g=1.82$~eV, the calculated binding energies of the $1s$, $2p_{\pm}$, and $2s$ states are approximately 126, 61, and 36~meV, respectively. These values are obtained from the numerical diagonalization of the internal exciton Hamiltonian. Higher excited states, such as $3s$, $3p_{\pm}$, and $3d_{\pm}$, are also included in the underlying exciton basis; however, they are not explicitly resolved in the spectrum shown here because their comparatively weak oscillator strengths result in substantially lower optical spectral weight.
\\

In [Fig.~\ref{fillingPL}(b)], we present the corresponding absorption map of the heterostructure around the light cone ($\mathbf{Q}=0$), calculated using the expression

\begin{equation}
    \alpha_{\nu}(\hbar\omega)\propto \sum_{l_i}{\big|\int\Psi_{\nu,l_i}(r=0, \textbf{R})d^2\textbf{R}}\big|^2_{\textbf{Q}=0}\frac{\gamma}{\pi[\hbar\omega-E_{\nu,l_i}]^2+\gamma^2},
\end{equation}

 where the moiré exciton eigenstates are expressed as
$\Psi_{\nu,i}(r,\mathbf{R})=\sum_{l_i} C_{\nu,l_i}\phi_{n\ell}(r=0)e^{-i\mathbf{g}_j\cdot\mathbf{R}}$,
with $C_{\nu,l_i}$ denoting the expansion coefficients obtained from the numerical diagonalization for the eigenstate indexed by $i$. Here, $\gamma=1$ meV is the phenomenological broadening parameter, and $E_{\nu,i}$ represents the corresponding moiré exciton eigenenergy.

As shown in [Fig.~\ref{fillingPL}(b)], the moiré-exciton resonances exhibit an approximately linear redshift with increasing filling factor $\nu$. This behavior is particularly pronounced for the $|2s;\mathbf{g}_j\rangle$ state and the lowest-energy $|1s;\mathbf{g}_j\rangle$ and $|2p_{\pm};\mathbf{g}_j\rangle$ states, with the optical response remaining dominated by the $|1s;\mathbf{g}_j\rangle$ exciton. For clarity, the oscillator strengths of the excitonic states with principal quantum number $n>1$ have been multiplied by a factor of ten.

The energy shift reaches its maximum at a characteristic filling factor of $\delta\nu\simeq20$, corresponding to a carrier density of approximately $4\times10^{12}~\mathrm{cm}^{-2}$. In particular, the lowest-energy $|2p_{\pm};\mathbf{g}_j\rangle$ state exhibits the largest filling-induced shift, reaching $\Delta E_m\simeq 82$ meV, as highlighted in [Fig ~\ref{fillingPL}(b)]. Here, $\Delta E_m$ is defined as the energy separation between the $|2s;\mathbf{g}_j\rangle$ state at charge neutrality ($\nu=0$) and the minimum energy of the lowest-energy $|2p_{\pm};\mathbf{g}_j\rangle$ branch reached upon increasing the filling factor.

\begin{table}[t]
\centering
\caption{Comparison with experiment and theoretical works.}
\label{tab:comparison}
\begin{tabular}{|c|c|c|c|c|c|c|}
\hline
 & \multicolumn{2}{|c|}{This work}
 & \multicolumn{2}{|c|}{Exp~\cite{Hu2023}}
 & \multicolumn{2}{|c}{theo~\cite{Lu2024}} \\
$\theta_t$(°) & $\Delta E_m$ & $\delta\nu$
         & $\Delta E_m$ & $\delta\nu$
         & $\Delta E_m$ & $\delta\nu$ \\
 & (meV) & & (meV) & & (meV) & \\
\hline
$0.6^\circ$ & 82 & 20 & 73 & 20 & 70 & 20 \\
$0.7^\circ$ & 56 & 15 & -- & -- & 67 & 15 \\
$0.8^\circ$ & 50& 13 & 53 & 15 & --& -- \\
$0.94^\circ$ & 33 & 11 & -- & -- & 45& 11\\
\hline
\end{tabular}
\end{table}

 Interestingly, the $|2p_{\pm};\mathbf{g}_j\rangle$ states acquire a finite optical oscillator strength at finite carrier doping. This doping-induced brightening originates from the hybridization of the $2p_\pm$ exciton states with higher-lying $s$-type excitonic states, in particular the $2s$ state. Microscopically, this mixing arises from the exciton COM coupling mediated by the periodic electrostatic moiré potential, as described by [Eqs.~(\ref{Eq.H}) and (\ref{kernal})]. The spatially periodic potential couples excitonic states with different internal angular-momentum character and thereby breaks the rotational symmetry of the $2p_\pm$ exciton state. As a result, the optically dark $2p_{\pm}$ states acquire an admixture of optically active $s$-type character, allowing otherwise dipole-forbidden transitions to acquire a finite oscillator strength.

 For a qualitative comparison with experiment, our theoretical results are consistent with the experimentally measured carrier-density-dependent reflectance contrast, $dR/R_0$, reported for the same heterostructure at low temperature ($T=1.7$~K), as shown in \cite{Hu2023}. In particular, the evolution of the excitonic resonances with increasing carrier density follows the same overall trend as observed experimentally.
\\

In [Fig~\ref{fillingPL}(c)], we extend our analysis to examine the dependence of the characteristic quantities $\Delta E_m$ and $\delta\nu$ on the moiré period $a_m$, and hence on the twist angle. We find that $\Delta E_m$ increases monotonically with increasing $a_m$, as indicated by the cyan curve. In contrast, the characteristic filling $\delta\nu$ exhibits an approximately linear dependence on $a_m$, as shown by the red curve. The calculated dependence is fitted by $\delta\nu=q_\nu a_m+\nu_0$ (dashed red curve), where $q_\nu$ = 0.93 $nm^{-1}$ denotes the slope and $\nu_0=-3.55$ is the intercept.
In [Fig \ref{fillingPL}(d)], we further investigate the dependence of $\Delta E_m$ on the interlayer separation $d$ between the $WSe_2$ monolayer and the adjacent tBG. A quasi-linear decrease of $\Delta E_m$ is obtained with increasing $d$.
\\

For a quantitative comparison with previous studies, we compare the calculated values of $\Delta E_m$ and $\delta\nu$ with the corresponding experimental results reported in \cite{Hu2023} and the theoretical predictions of \cite{Lu2024}, as summarized in [Table~(\ref{tab:comparison})].

 In [Figs.~\ref{fillingPL2}(a--c)], we investigate the effect of the  twist angle  of the tBG on the optical spectra for $\theta_t=0.54^\circ$, $0.70^\circ$, and $0.80^\circ$, respectively. At the smallest twist angle, $\theta_t=0.54^\circ$, a pronounced avoided crossing is observed between the lower-energy $|2p_{\pm};\mathbf{g}_j\rangle$ branch and the $|1s;\mathbf{g}_j\rangle$ state. As the twist angle increases, the energy separation between the lowest-energy excitonic branch and the higher-lying excited states becomes progressively larger, indicating a systematic modification of the excitonic level structure with the moiré period.
To further elucidate the origin of this behavior, we analyze the corresponding exciton density distributions for the eigenstates indexed from 8 to 12 at a fixed twist angle of $\theta_t=0.45^\circ$ and filling factor $\nu=10$ [see Figs.~\ref{fillingPL2}(i--m)]. These density distributions reveal the distinct spatial character and deformation of the relevant excitonic states, providing further insight into the hybridization and level evolution observed in the optical spectra.
\\

It is worth noting that several effects that can contribute to the fine structure of the excitonic Rydberg states are not included in our present model. First, we neglect the exciton fine structure associated with the Berry curvature of the underlying Bloch states, which can lift the degeneracy between the $2p_{+}$ and $2p_{-}$ Rydberg states. We also neglect the electron--hole exchange interaction associated with the twofold valley pseudospin degeneracy at the $K$ and $K'$ valleys. In addition, although the screened electron--hole Coulomb interaction naturally lifts the degeneracy between the $2s$ and $2p$ states, the magnitude of this splitting depends sensitively on the dielectric environment and screening length. For hBN-encapsulated $WSe_2$, the $2s$--$2p$ separation has been reported to be only a few meV \cite{Kim2025,Lu2024}, substantially smaller than the corresponding separation expected for a more strongly non-hydrogenic Coulomb potential. This relatively small splitting can be understood from the long-wavelength limit of the Rytova--Keldysh interaction, $q\ell_0\ll1$ (or equivalently $r\gg\ell_0$), where the screened interaction approaches the 2D Coulomb form, $U_{\mathrm{sc}}(q)\sim1/(\epsilon q)$, thereby recovering an approximately hydrogenic Rydberg spectrum.
Furthermore, the additional screening of the electron--hole interaction induced by the nearby doped tBG layer is not explicitly included in the present treatment. A quantitative description of this contribution would require a microscopic calculation of the dynamically screened interaction, for example within the random-phase approximation (RPA), including the electronic response of the tBG layer.  
In addition, the experimentally observed dependence of the $2s$ exciton resonance on integer filling factors near the magic angle \cite{Hu2023,Hu2025}, characterized by a characteristic sawtooth-like structure, is not reproduced in the present calculation. Capturing this feature would likely require a more microscopic treatment of the carrier-density dependence, including the detailed electronic structure and filling-dependent density of states of the tBG. These extensions would provide a more complete description of the many-body and filling-dependent effects in the $WSe_2/tBG$ heterostructure.

Consequently, the present model should be regarded as an effective description that captures the dominant doping-dependent evolution of the moiré exciton spectrum while neglecting these additional fine-structure and many-body screening effects. Within this level of approximation, the calculated spectral evolution remains qualitatively consistent with the experimentally observed carrier density dependent excitonic features reported in ~\cite{Hu2023}.

\section{Conclusions}\label{sec4}

 In this work, we have investigated the effect of electrostatic doping on the hBN-encapsulated $WSe_2/tBG$ heterostructure. The doped tBG  generates an effective mean-field periodic electrostatic potential that couples to the electron and hole constituents of an exciton in the adjacent $WSe_2$ monolayer, thereby modifying its optical response. We have developed a microscopic model of this coupled system and investigated the evolution of the excitonic optical spectrum as a function of experimentally relevant parameters, including the tBG twist angle and the interlayer separation between the $WSe_2$ monolayer and tBG.
A central result of this work is that the interplay between electrostatic doping and the interlayer degree of freedom, particularly for twist angles below the magic angle, provides a mechanism for modifying and optically activating the Rydberg exciton states through orbital hybridization. In particular, the coupling between different excitonic orbital characters can transfer oscillator strength to otherwise weakly or optically inactive $2p_{\pm}$ states, enabling their signatures to emerge in the optical spectrum. The calculated doping-dependent evolution of the excitonic resonances is qualitatively consistent with previous experimental observations and theoretical predictions. These results highlight the sensitivity of moiré exciton Rydberg states to the electrostatic environment and interlayer geometry, and suggest that doped $WSe_2/tBG$ heterostructures may provide a platform for exploring optically tunable excitonic states with potential relevance to nanoscale optical sensing.
 \\

\textbf{Acknowledgement.} H.E.H acknowledges support by the  Laboratory of Materials Physics, Faculty of Sciences of Bizerte, University of Carthage.
\\

\textbf{ Data availability:} 
The datasets used and/or anlyzed during the current study are available from the corresponding author on resonable request.


\end{document}